\documentclass[12pt]{article}

\usepackage{graphics}
\usepackage{epsfig}
\usepackage{a4wide}
\usepackage{latexsym}
\usepackage{amssymb}
\usepackage{amsfonts}

\begin{document}

\title{Green's function probe of a static granular piling}

\author{Guillaume Reydellet, Eric Cl\'{e}ment\\ \\
Laboratoire des Milieux D\'{e}sordonn\'{e}s et H\'{e}t\'{e}rog\`{e}nes \\ 
UMR7603 - Universit\'{e} Pierre et Marie Curie - Bo\^{\i }te 86\\ 
4, Place Jussieu, 75005\ Paris
}

\setcounter{page}{0}
\maketitle

\vspace{1cm}

\begin{abstract}
We present an experiment which aim is to investigate the mechanical 
properties of a static granular assembly. The piling is an horizontal 3D 
granular layer confined in a box, we apply a localized extra force at the 
surface and the spatial distribution of stresses at the bottom is obtained 
(the mechanical Green's function). For different types of granular media, we 
observe a linear pressure response which profile shows one peak centered at 
the vertical of the point of application. The peak's width increases 
linearly when increasing the depth. This green function seems to be in -at 
least- qualitative agreement with predictions of elastic theory.
\end{abstract}

\vspace{1cm}

\thispagestyle{empty}


\newpage

Understanding the exact mechanical status of static or quasi- static 
granular assemblies is still an open and debated issue\cite{PDM98},\cite 
{degennes99}. So far, there is no consensus on how to express correctly the 
stress distributions in a granular piling under various boundary conditions. 
Traditional approaches of soil mechanics typically use\ elasto-plastic 
modeling for granular materials\cite{Feda82} and constitutive relations are 
obtained empirically from standart triaxial tests. In this picture, for 
small deformations, an elastic-like behavior is assumed and a set of 
elliptic partial differential equations (P.D.E.) is then used to get the 
stress/strain distributions. For larger strains, the Coulomb plasticity 
theory is adapted to model granular flows, which involves hyperbolic 
(propagative) PDE's in regions experiencing yield \cite{Nedderman92}. At the 
granular scale, several recent experiments and simulations have evidenced 
the presence of a rather large distribution of contact forces\cite{Mueth98} 
between the grains as well as force chains \cite{Radjai96},\cite{Howell99} 
spanning a volume of 10 to15 grains sizes. Hence it is clear that a rigorous 
passage from a microscopic to a macroscopic mechanical description that 
would include this mesoscopic disorder is an arduous task and a challenging 
problem of statistical physics. Consequently, these studies have triggered 
alternative theoretical approaches. One of them is based on a scalar 
stochastic modeling\cite{Liu95} for the contact force redistributions. In 
the large scale limit, this vision provides a diffusive-like picture 
(parabolic equation ) for the stress transmission properties. Another 
approach incorporates the vectorial and propagative character of the contact 
forces between the grains\cite{Bouchaud95},\cite{Edwards96}. This picture, 
when extended to the continuum limit, predicts simple relations between the 
components of the stress tensor and implies hyperbolic (i.e. propagative) 
PDEs for the stress fields\cite{Bouchaud95}, \cite{Cates98}. A\ recent 
framework bridges the two last approaches \cite{Krenkre98}.\  
 
Several reproducible experiments were recently performed on a sand-heap (see  
\cite{Vanel99a} and references therein) and on a granular column \cite 
{Vanel99b}. It was shown that an hyperbolic modeling (for example the OSL 
model\cite{Cates98}) is indeed able to reproduce some of the observed 
phenomenology. But it is worth noting that the parameters entering in these 
hyperbolic models are so far phenomenological constants, calculated a 
posteriori and therefore, do not provide us with really predictive 
statements. Furthermore, the agreement between the available experiments and 
the hyperbolic models does not rule out the pertinence of elliptic models  
\cite{Cantelaube98},\cite{Savage98}. 
 
Here, we present an experiment probing the response of static granular 
assemblies to a local stress perturbation (a Green's function). This is 
probably the most basic experiment allowing a precise discrimination between 
the different approaches and which should reveal the real mechanical nature 
of static granular assemblies\cite{degennes99}. We operate on an horizontal 
3D granular assembly confined in a box and the spatial distribution of 
stresses at the bottom is monitored which provides an important piece of 
evidence in order to inform this currently debated issue. 
 
Already at the most basic level, measuring meaningful stresses in granular 
assemblies is a non trivial question. Generally, experimentalists measure 
stresses by calibration of devices that are deformed or displaced as a 
result of a local force distribution on the probe surface. They are 
confronted to three fundamental problems, 
 
i) the response of the probe depends on the history of the preparation as 
evidenced on sand heap\cite{Vanel99a} (which might be a physically relevant 
issue), 
 
ii) the probe surface has a limited number of contacts with the granular 
medium which is at the origin of an inherent fluctuations scale (which 
importance should decrease when the probe size increases), 
 
iii) the physical characteristics of the probe itself may have an influence 
on the measurements. 
 
In this last situation, large deformations of the device may change 
drastically the local force distribution creating arching effects around the 
probe. The stress probe we use here is made of a thin metallic membrane of 
thickness $e=100\mu m$ welded on a cylinder. The deformation of the membrane 
(less than $1\mu m$) is monitored using a sensitive capacitive technique. To 
avoid the formation of a vault around the probe and break the history 
dependence when building the sand layer, the pile is slightly vibrated after 
each height increase and the sand is randomly poured by hand layer by layer. 
This procedure produces some fluctuations from one step to the other but in 
the average, we checked that, locally, the hydrostatic pressure relation $%
P=\rho gh$ \ is well recovered (with a precision of $5\%$, which is the 
typical uncertainty on the packing fraction determination). For all the 
experiments we describe here, we use this procedure to prepare the sand-pile. 
 
Probing the response of a granular static piling to a localized perturbation 
is a priori a difficult issue. It is of common experience that it is quite 
easy to drill a hole in a sandy surface when pushing it weakly with the tip 
of a finger. Therefore, since here the perturbing stress must not create 
plastic reorganization of the grains, only weak perturbations must be 
applied. Consequently, the detection of the response signal is likely to be 
rapidly hidden within the noise when depth increases. The typical pressure 
we apply at the surface of the pile is achieved by a piston $P$ (see fig.1) 
of mass $M=5g$ with a surface $A=1cm^{2}$ ($5.10^{2}$ $Pa$). In order to 
increase the signal/noise ratio, we use a lock-in detection technique. More 
precisely, a local stress modulation is achieved using a periodic magnetic 
field created by an electric current in a coil surrounding the piston (see $%
C $ on fig1). In the piston, a permanent magnet is inserted \ and the coil 
current is driven by a low frequency generator ($G$). Therefore, the 
modulation of the magnetic field in the coil creates a force modulation on 
the piston.. The signal of the stress probe $P1$ at the bottom of the piling 
is then directed to the lock-in amplifier $L$ synchronized by the generator 
exciting the source. Note that a sensitive displacement probe $P2$ monitors 
the piston position to check that no plastic yield occurs during the data 
collection. The relative horizontal position $x$ between the piston and the 
probe $P1$ is varied. We operate in the low frequency limit such that we are 
basically probing the static properties of the granular piling. The applied 
extra modulated force is driven at $f=\omega /2\pi =80Hz$ and we checked 
that the exact choice of this frequency modulation does not matter in this 
limit (between$10$ and $120Hz$). We also verified that within a reasonable 
time scale (several tenths of minutes), we do not observe slow variation of 
the response. Due to the finite sizes of the piston and of the probe, the 
signal hence obtained, is the convolution of the mechanical response 
function (the Green's function) by the width of the source and the width of 
the probe. We measured the intrinsic experimental width $W_{o}=10mm$ and we 
found the convolution effects to be negligible as soon as $h>3cm$ 
(corresponding to $3$ times the probe diameter). 
 
Here, we report experiments on there granular media with different size $d$ 
and shape. We use $d\simeq 1mm$ ''aquarium sand'' and $d\simeq 300\mu m$ 
''Fontainebleau sand''. The grain shape is rather rough and the size 
polydispersity is around $50\%$. We also use monodisperse glass beads 
(diameter $d=1.5mm$). We tested that, in the limit where no ''sinking'' of 
the piston inside the pile is observed, the response is linear in the value 
of the imposed stress. We also found that the value of the slope relating 
the applied force to the observed stress may depend strongly on granular 
configurations around the probe. This ''realization dependence'' causes 
difficulties to calibrate precisely the probe at $80Hz$ on a granular 
column. This is evidenced on the inset of fig.1 where we display the stress 
measurement for three different piles obtained in the same conditions. The 
response amplitude after detection by the lock-in amplifier at $x=0$ is 
plotted as a function of the force modulation amplitude $F$ . Note that the 
force values stem from a calibration on a static water column and not on the 
granular pile response at $80Hz$ . The frequency of the signal is small 
enough such that we have no significant phase shift between the force and 
the detected stress. Moving the point of application of the force, we change 
the horizontal distance $x$ between the piston and the probe and plot the 
pressure profile $\sigma _{zz}(x)$ for a given depth $h$ of sand. 
 
The response function $P(x)$ shows one single peak centered at the vertical 
of the point of application of the force, we did not observe the two 
separated bumps as predicted by hyperbolic models\cite{Claudin98},\cite 
{degennes99}, even when increasing the depth up to $10cm$ which corresponds 
to the limit of our detection possibilities. On Fig 2a, we display the 
response functions rescaled by an amount $P^{\ast }$ so that they present 
the same area under the curve (i.e. constant applied force). Importantly, we 
checked that for three different probes situated at three difference places 
on the bottom plate, this procedure provides us with the same stress 
distribution. On fig.2a, we display three response functions $P(x)$ obtained 
at different depths $h$ for the $d=1mm$ sand. We find that for all granular 
media studied, the width at half amplitude $w$ increases linearly when 
increasing the depth (whenever $h>W_{0}$). The slope is independent of the 
material used (see fig 2b). This is in qualitative agreement with the 
prediction of elastic theory.\ Here we provide a comparison with the only 
known exact solution in $3D$ obtained for an infinite half space as computed 
by Boussinesq and Cerruti last century \cite{Johnson85}: 
 
\[ 
\sigma _{zz}=%
{\displaystyle{-3F \over 2\pi }}%
.%
{\displaystyle{z^{3} \over (x^{2}+z^{2})^{\frac{5}{2}}}}%
\] 
where $F$ is the applied force. On fig 3 we plot the response functions at 
different depths $h$ such that the response $P(x)$ is rescaled by $z^{2}$ 
and the horizontal axis is $x^{\ast }=x/h$. We see that for all the granular 
material tested, the curves are collapsing onto the same function. The 
response is clearly sharper than the elastic lorentzian prediction obtained 
for a semi-infinite medium. We get $w/h=0.94\pm 0.05$ instead of $%
w/h=1.13... $ (i.e. $20\%$ sharper). One problem here is to account 
correctly for the bottom boundary condition which is not a trivial issue 
since many choices can be done and its requires an adapted elastic code. 
Nevertheless, we are aware of an analytical calculation performed on the 
same problem in $2D$ and in $3D$ by Claudin et al.\cite{Claudin00}. Rough 
and smooth boundary conditions are used. Boths calculation shows a clear 
sharpening of the response of about $20\%$ in $2D$ and $6\%$ in $3D$ when 
compared to the exact semi-infinite solution in 2D. Thus the issue is subtle 
and requires precise finite element elastic calculations. We leave the 
discussion for future investigations. Moreover, it is important to notice 
that the linear broadening is not consistent with any leading parabolic 
behavior \cite{Liu95} \cite{Krenkre98} on large scales where broadening 
increasing as a square root of depth would be expected. Therefore,our 
results clearly contradicts the claim of generic parabolic behavior 
extracted from recent experiments \cite{Rajchenbach00} done a quite specific 
granular assembly and obtained at a very small scale. 
 
In conclusion, we present the first experimental determination of the 
horizontal stress distribution in response to a localized stress 
solicitation (i.e. the mechanical response function or Green's function) in 
a granular piling. The stress solicitations are made along the vertical axis 
and the spatial distribution of pressures are measured at the bottom of the 
pile. The different pililing tested were very disordered in terms of size, 
polydispersity and friction between the grains. The piling procedure we use 
avoids as much as possible, preparation memory effects. In such a case, the 
Green's function is consistent with predictions of elasticity since it does 
not exhibit the two side bumps as the hyperbolic modelling would. We also 
find a linear dependence of the half-height enlargement width depth. This 
last result rules out parabolic modelling of disordered granular assemblies 
(on a large scale, at least). Nevertheless, we do not find so far a complete 
quantitative agreement with the exact result of\ isotropic elasticity in an 
infinite half space. An open question is still to understand how and whether 
this difference could be captured when considering explicitely boundary 
conditions imposed by the bottom plate. This issue is left out for future 
report. On the other hand, it would be also quite interesting to know 
whether predictions of hyperbolic models would be drastically changed when 
including strong disorder in the theory. In further experiments, we will 
study specifically the role of disorder and\ texture in relation with their 
eventual influence on the mechanical Green' function.

We thank P.Claudin, J.-P. Bouchaud and Prof. R.P.Behringer for many fruitful 
interactions.\ We acknowledge the financial support of the grant PICS-CNRS $%
\#563.$


\newpage

\begin{center}
\includegraphics[width=\linewidth]{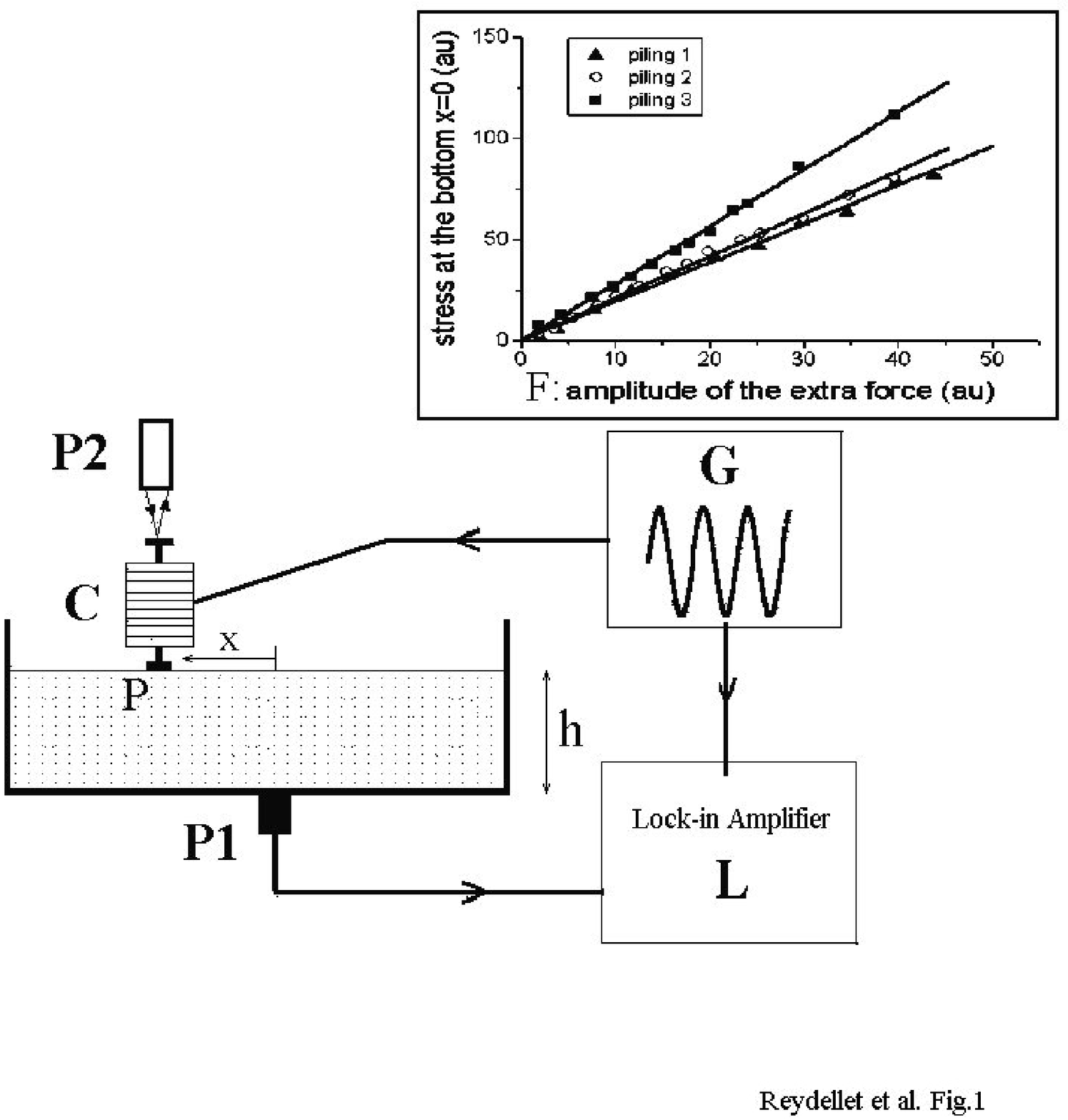}
\end{center}

Sketch of the experimental display : $G$ low frequency generator, $C$ 
electrical coil, $P$ piston , $P1$ stress probe, $P2$ displacement probe, $L$ 
lock-in amplifier (see text for a detailed description).\ Inset : test of 
the response linearity for three independent experiments on the same probe $%
P2$. The axes (stress versus applied force) are labeled in arbitrary units.

\newpage

\begin{center}
\includegraphics[height=19cm]{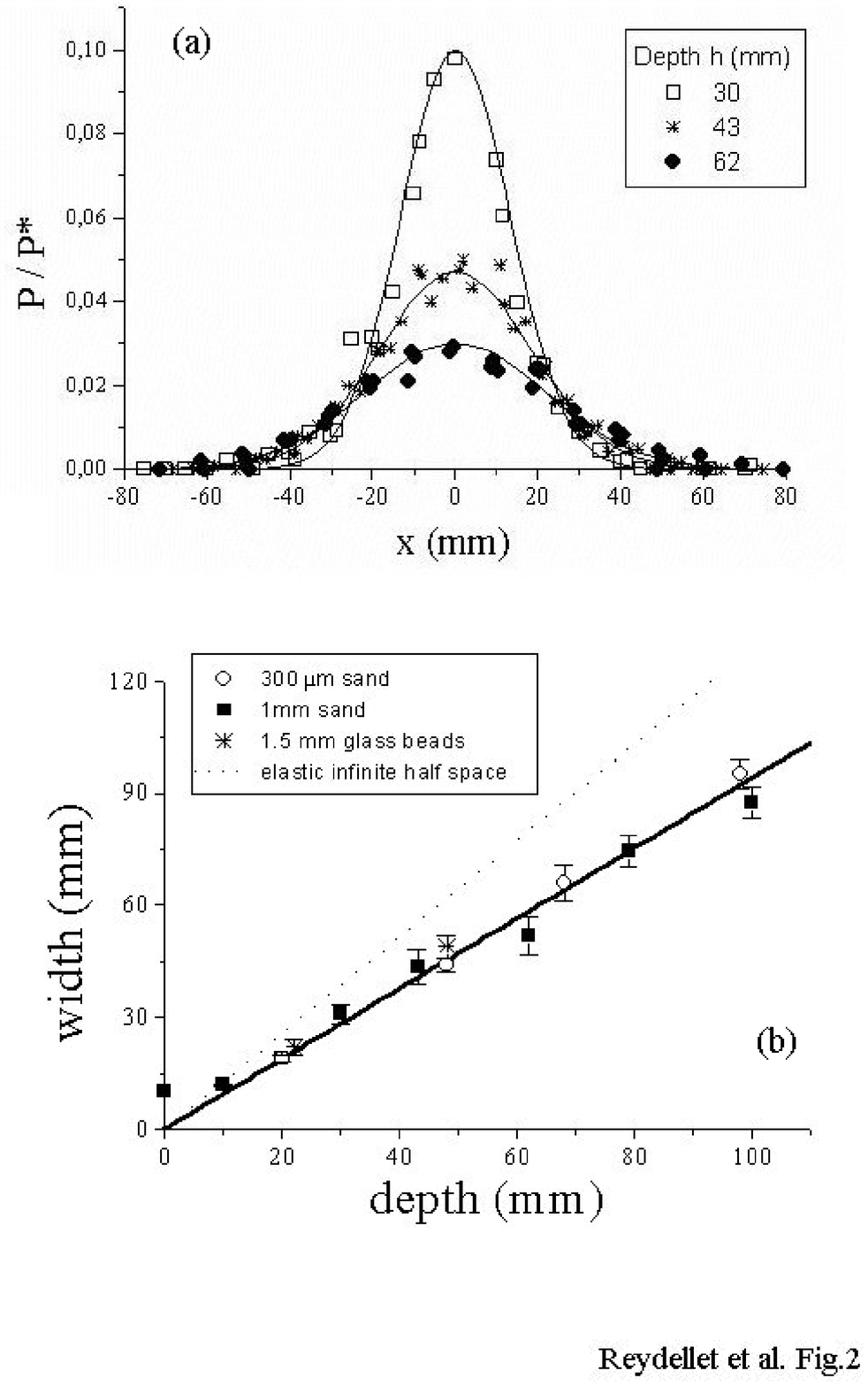}
\end{center}

Horizontal stress distribution in response to a localized 
solicitation (Green's function). Fig. 2a : Green's function $P(x)=\sigma 
_{zz}(x)/P^{\ast }$ measured at three different depths for $d=1mm$ sand. See 
text for definition of the rescaling factor $P^{\ast }$. Fig2b : half 
amplitude width $W$ of the response function as a function of depth $h$ for 
three different granular materials (see legend). The straight line is the 
best linear fit : $W=0.94h$.

\newpage

\begin{center}
\includegraphics[width=\linewidth]{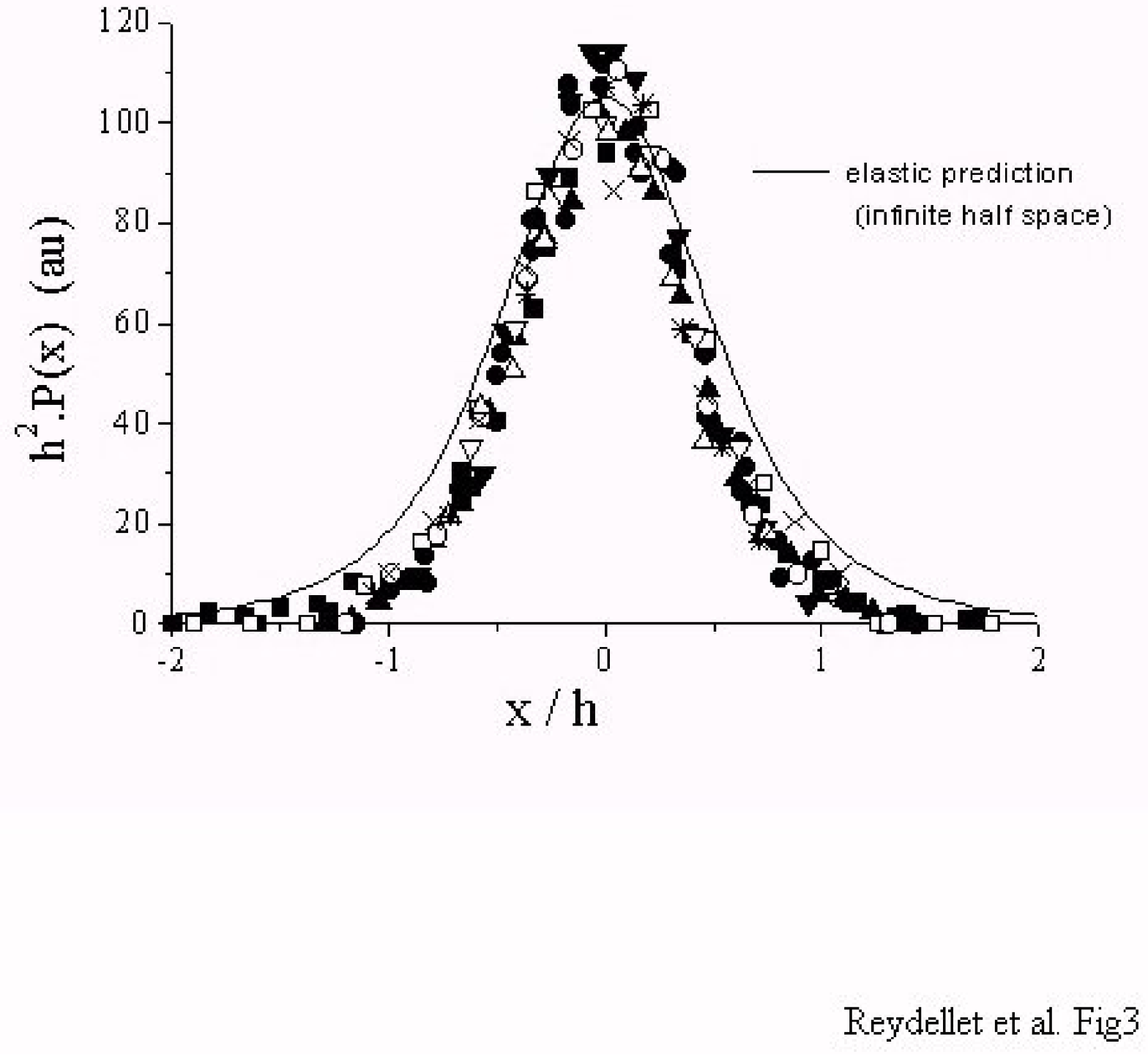}
\end{center}

Rescaled Green's function $h^{2}P(x)$ as a function of the rescaled 
horizontal axis $x/h$, for different depths and different types of granular 
materials. Aquarium sand : $d=300\mu m$ (\ $h=30mm(\blacksquare 
),62mm(\bullet ),79mm(\blacktriangle ),100mm(\blacktriangledown )$); 
Fontainebleau sand : $d=1mm$ ($h=19mm(\square ),48mm(\bigcirc 
),68mm(\bigtriangleup ),97mm(\bigtriangledown )$); glass beads : $d=1mm$ (\ $%
h=28mm(\divideontimes ),48mm(\times )$). The straight line is the 
theoretical response of an elastic infinite half-space.

\end{document}